\documentclass[10pt,a4paper,twocolumn,aps,prl]{revtex4}
\usepackage{amsmath,bm}
\usepackage[dvips]{graphicx,color}
\usepackage[latin1]{inputenc}
\usepackage[T1]{fontenc}
\setcitestyle{super}
\usepackage{latexsym}
\usepackage{times}
\usepackage{float}

\begin{document}

\title{Einstein-Podolsky-Rosen paradox in twin images}

\author{Paul-Antoine Moreau, Fabrice Devaux, and Eric Lantz}
\affiliation{Département d'Optique, Institut FEMTO-ST, Université de Franche-Comté, CNRS, Besançon, France}

\date{\today}

\maketitle

\noindent\textbf{ Spatially entangled twin photons provide both promising resources for modern quantum information protocols, because of the high dimensionality of transverse entanglement\cite{Howland2013, Dixon2012}, and a test of the Einstein-Podolsky-Rosen (EPR) paradox\cite{Einstein1935} in its original form of position versus impulsion. Usually, photons in temporal coincidence are selected and their positions recorded, resulting in a priori assumptions on their spatio-temporal behavior \cite{Howell2004}. Here, we record on two separate electron-multiplying charge coupled devices (EMCCD) cameras twin images of the entire flux of spontaneous down-conversion. This ensures a strict equivalence between the subsystems corresponding to the detection of either position (image or near-field plane) or momentum (Fourier or far-field plane)\cite{Reid2009}. We report the highest degree of paradox ever reported and show that this degree corresponds to the number of independent degrees of freedom \cite{Law2004,Exter2006}\ or resolution cells \cite {Devaux95}, of the images.}

In 1935, Einstein, Podolsky and Rosen (EPR) showed that quantum mechanics predicts that entangled particles could have both perfectly correlated positions and momenta, in contradiction with the so-called \textit{local realism} where two distant particles should be treated as two different systems. Though the original intention of EPR was to show that quantum mechanics is not complete, the standard present view is that entangled particles do experience nonlocal correlations \cite{Bell1964,Aspect1981}. It can be shown that the spatial extent of these correlations corresponds to the size of a spatial unit of information, or mode, offering the possibility of detecting high dimensional entanglement in an image with a sufficient number of resolution cells\cite{Dixon2012,Howell2004}. However, in most experiments the use of single photon detectors and coincidence counting leads to the detection of a very few part of selected photons, generating a sampling loophole in fundamental demonstrations. High sensitivity array detectors have been used outside the single photon-counting regime in order to witness the quantum feature of light, showing the possibility of achieving larger signal-to-noise ratio than in classical imaging \cite{Brida2010,Ottavia04}. However, the EPR paradox is intimately connected to the particle character of light and its detection should involve single photon imaging, possible either with intensified charge coupled devices (ICCD) or, more recently, EMCCDs \cite{Lantz2008}. ICCDs exhibit a lower noise but have also a lower quantum efficiency than EMCCDs and a more extended spatial impulse response. ICCD are therefore convenient to isolate pairs of entangled photons \cite{Oemrawsingh2002}, as shown in a recent experiment: an ICCD triggered by a single photon detector was used to detect heralded photons in various spatial modes\cite{Fickler2013}.

On the other hand,  because of their higher quantum efficiency EMCCDs allow efficient detection of quantum correlations in images, as demonstrated some years ago by measuring sub-shot-noise correlations in far-field images of spontaneous parametric down-conversion (SPDC)\cite{Blanchet2008,Blanchet2010}. More recently, two experiments intended to achieve the demonstration of an EPR paradox in couples of near field and far-field images recorded with in an EMCCD. The first experiment, in our group, involved the detection of twin images on a single camera, by separating in the near-field the cross-polarized photons with a polarizing beam-splitter, inducing some overlap of the near-field images and a  rather small resolution in the far-field because of walk-off. The results exhibited a low degree of paradox, far from the theoretical values, though highly significant and in accordance with the full-field requirements\cite{Moreau2012}.  The second experiment\cite{Edgar2012} exhibited also both near field (position) and far field (momentum) correlations, with a much lower product of the spatial extents. However only one beam of particles was involved, because of type-I phase matching, making the results questionable if intended as a demonstration of an EPR paradox. Moreover, the results were obtained for only one dimension because of smearing effects and detection of correlations between adjacent pixels. \begin{figure*}[t]
  \centerline{\includegraphics[width=18cm]{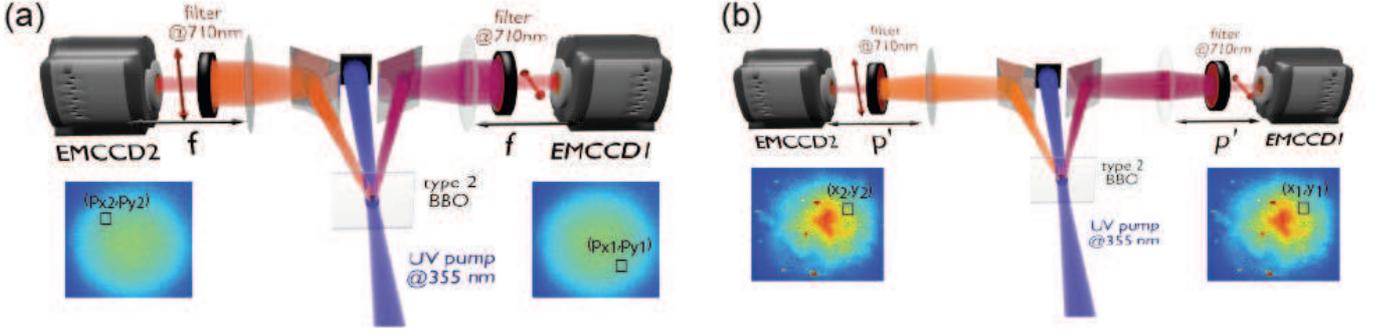}}
  \caption{\textbf{Experimental setups used to imaging correlations.} (\textbf{a}), measurement of momentum correlations with the cameras in the focal plane. Inserts: sums of 700 far-field images; $px_{1}=-px_{2},py_{1}=-py_{2}$ are the coordinates of twin pixels. (\textbf{b}), cameras in the crystal image plane and sums of 700 near-field images with twin pixels in $x_{1}=x_{2},y_{1}=y_{2}$.}
\label{fig1}
\end{figure*}
\begin{figure*}[t]
  \centerline{\includegraphics[width=16cm]{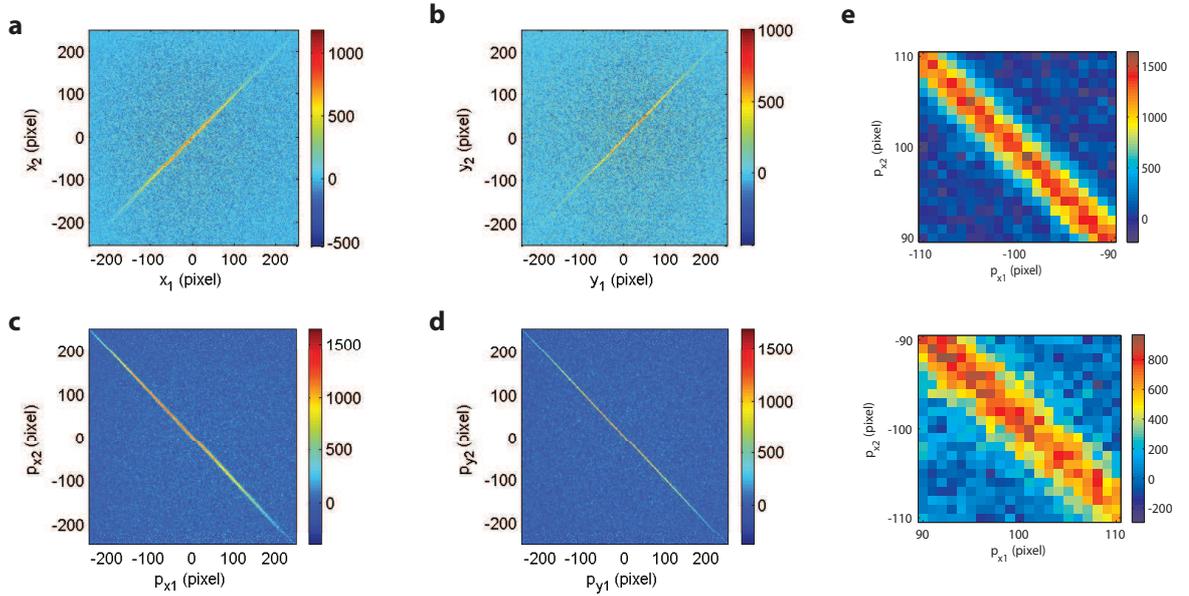}}
  \caption{\textbf{Joint probabilities versus the transverse spatial coordinates} Color scales are expressed in coincidence counts over 35000 pairs of images, corrected from the mean corresponding to accidental coincidences. (\textbf{a},\textbf{b}): near-field. (\textbf{c},\textbf{d}): far-field. (\textbf{e}): In the far field, the correlations arise in a coherence area that is larger for momenta the most distant from the pump direction, i.e. for the largest values of $x_{1}$, x being the direction along which the two fluorescence beams are separated from the walk-off. This observation is explained by the finite bandwidth of the interference filter\cite{Devaux2012}, that leads to the detection of twin photons out of the degeneracy, $\lambda_1\neq\lambda_2$.
  }
  \label{fig3}
  \end{figure*}

In the present experiment, the use of two cameras allows a separation of the twin images without any further optical component, thanks to walk-off, and a perfect identity of the subsystems but the position of the imaging systems, composed on each arm of a lens and a camera. Before describing our experimental results, let us recall that an EPR paradox arises when correlations violate an inequality corresponding to the Heisenberg uncertainty principle if applied to a single particle 1 or 2, but expressed in terms of conditional variances\cite{Reid1989,Reid2009}:
\begin{equation}
\left\langle\Delta^2(\rho_1-\rho_2)\right\rangle\left\langle\Delta^2(p_{1}+p_{2})\right\rangle\geq\frac{\hbar^2}{4}\label{EPR}
\end{equation}
where $\rho_i$ is the transverse position of photon $i$ ($i=1,2$) at the center of the crystal and $p_i$ its transverse momentum. In order to make the demonstration consistent, the statistical average made to estimate the variances should be evaluated on the same system in the near and the far field. By using two EMCCD cameras that detect photons in the whole SPDC field, we ensured this consistency.
By approximating the phase matching function of SPDC to a Gaussian, the wave function of the biphoton can be written  \cite{Tasca2009}:
\begin{equation}
\Psi(\boldsymbol{\rho}_1,\boldsymbol{\rho}_2)=N\ exp\left(-\frac{\left|\boldsymbol{\rho}_1+\boldsymbol{\rho}_2\right|^2}{4\sigma_p^2}\right)
exp\left(-\frac{\left|\boldsymbol{\rho}_1-\boldsymbol{\rho}_2\right|^2}{4\sigma_{\phi}^2}\right)\label{eq1}
\end{equation}
\begin{equation}
\Psi(\boldsymbol{p}_1,\boldsymbol{p}_2)=\frac{1}{N\pi^2}\ exp\left(-\sigma_P^2\frac{\left|\boldsymbol{p}_1+\boldsymbol{p}_2\right|^2}{4\hbar^2}\right)exp\left(-\sigma_{\phi}^2\frac{\left|\boldsymbol{p}_1-\boldsymbol{p}_2\right|^2}{4\hbar^2}\right)\label{eq2}
\end{equation}
\begin{figure*}[t]
  \centerline{\includegraphics[width=16cm]{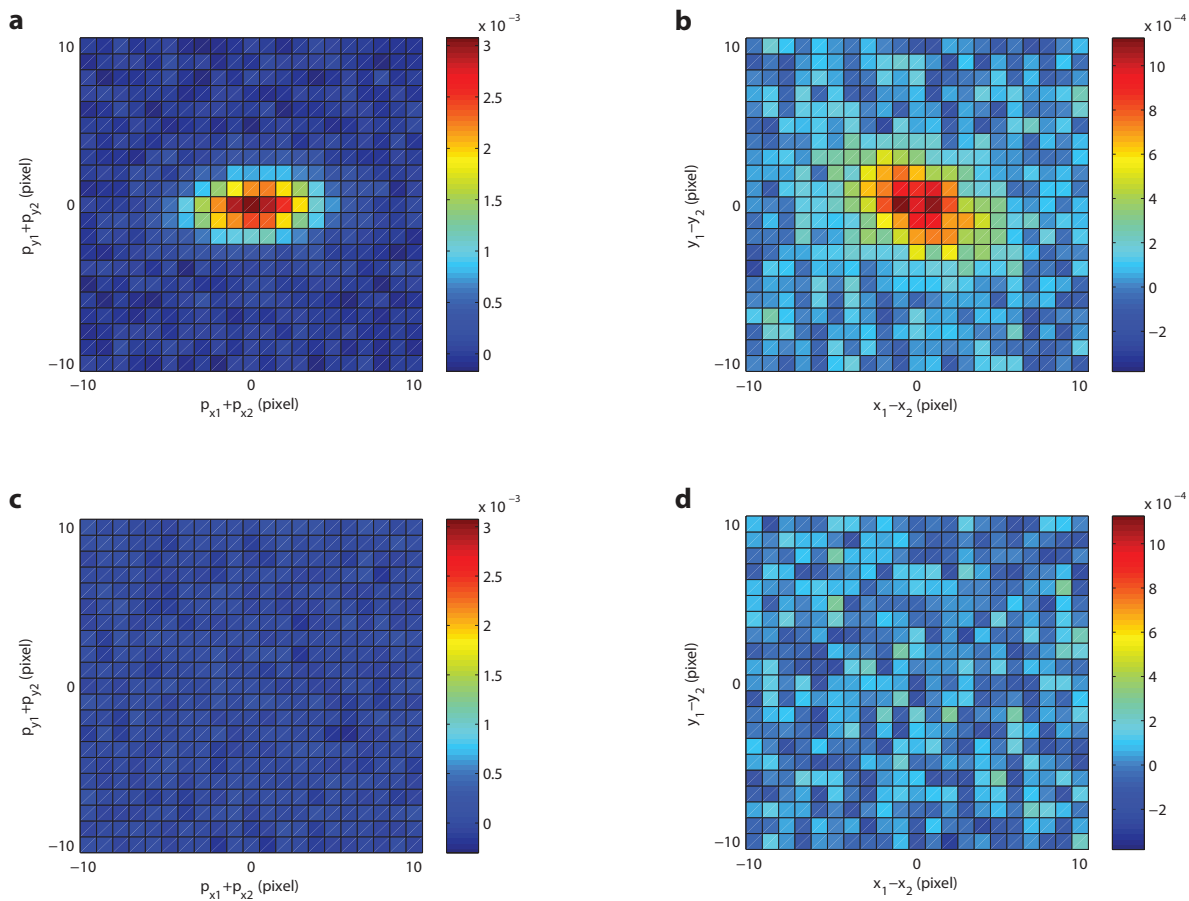}}
  \caption{\textbf{Normalized cross-correlation functions in position and momentum :} The cross-correlation is calculated over 700 images in the far-field (\textbf{a},\textbf{c}) and image plane (\textbf{b},\textbf{d}). In (\textbf{c}) and (\textbf{d}) are presented cross-correlation of images that do not share any pump pulses.}
\label{fig4}
\end{figure*}
where N is a normalization constant, $\boldsymbol{\rho}_i=(x_i,y_i)$, $\boldsymbol{p}_i=(p_{xi},p_{yi})$, $\sigma_P$ the standard deviation of the gaussian pump beam, and $\sigma_{\phi}$ the standard deviation, defined in the near-field, of the Fourier transform of the phase matching function defined in the far-field. In our experimental conditions where $\sigma_P>>\sigma_{\phi}$, these equations show that the product of conditional variances is equal to:
\begin{equation}
\left\langle\Delta^2(\rho_1-\rho_2)\right\rangle\left\langle\Delta^2(p_{1}+p_{2})\right\rangle=
\hbar^2\frac{\sigma_{\phi}^2}{\sigma_p^2}=\frac{\hbar^2}{4V}\label{K}
\end{equation}

where V is defined by this equation as the degree of paradox. Using results of Law and Eberly\cite{Law2004}, it can be shown that V is also the Schmidt number of the entanglement, i.e. the whole dimensionality of the biphoton in the two-dimensional transverse space supposed isotropic. For an unidimensional system, V becomes the square of the Schmidt number\cite{Fedorov2009}.

The experimental setup is shown in Fig.\ref{fig1}. Pump pulses at 355 nm provided by a 27 mW laser illuminated a $0.6$-mm long $\beta$ barium borate (BBO) nonlinear crystal cut for type-II phase matching. The signal and idler photons were separated by means of two mirrors and sent to two independent imaging systems. The far-field image of the SPDC was formed on the EMCCDs placed in the focal plane of two 120-mm lenses, Fig.\ref{fig1}a. In the near-field configuration, Fig.\ref{fig1}b, the plane of the BBO crystal was imaged on the EMCCDs with a transversal magnification $M=2.47\pm0.01$. Note that only the positions of the lenses and cameras are different in the two configurations. The back-illuminated EMCCD cameras (Andor iXon3) have a quantum efficiency greater than $90\%$ in the visible range. The detector area is formed by $512\times512$ pixels, with a pixel size of $s_{pix} = 16\times16\mu m^2$. We used a readout rate of 10 MHz at 14 bits, and the cameras were cooled to $-100^{\circ}$C. An image corresponds to the summation of 100 laser shots, i.e an exposure time of 0.1s and a dead-time between two successive images of about the same value, in order to allow a perfect synchronization between both cameras.  Measurements were performed for a crystal orientation corresponding to noncritical phase matching at degeneracy, i.e., collinear orientation of the signal and idler Poynting vectors in the crystal\cite{Lantz2000}. Photon pairs emitted around the degeneracy were selected by means of narrow-band interference filters centered at 710 nm ($\Delta\lambda = 4 nm$). The photon-counting regime was ensured by adjusting the exposure time in such a way that the mean fluency of SPDC was between 0.1 and 0.2 photon per pixel in order to minimize the whole number of false detections\cite{Lantz2008}. The mean number of photons per spatiotemporal mode was less than $10^{-3}$, in good agreement with the hypothesis of pure spontaneous parametric down-conversion, without any stimulated amplification. Following a method previously reported\cite{Lantz2008}, we applied a thresholding procedure on the image to convert the gray scales into binary values that correspond to 0 or 1 photon.

The conditional probability distributions calculated using 35000 images are shown in Fig.\ref{fig3}. The correlation profiles agree with the theoretical expectations (\ref{eq1}) and (\ref{eq2}) with $\sigma_p>>\sigma_{\phi}$. 

We have shown\cite{Devaux2012} that the conditional variances $\langle\Delta^2(\rho_1-\rho_2)\rangle$ and $\langle\Delta^2(p_1+p_2)\rangle$ correspond to the width of the normalized cross-correlation of photo-detection images. The experimental values obtained by fitting the normalized cross-correlations presented in Fig.\ref{fig4} are reported in Tab.\ref{tab}, for the two orthogonal directions of the transverse plane x and y.
\begin{figure*}[t]
  \centerline{\includegraphics[width=17cm]{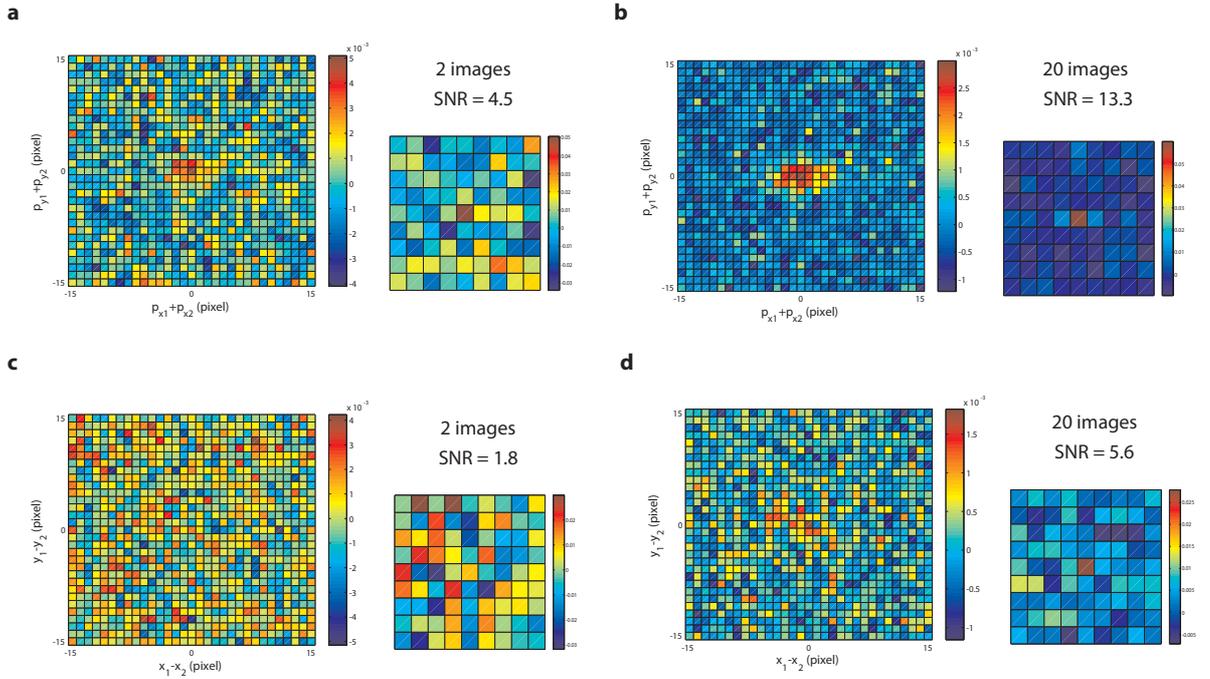}}
  \caption{\textbf{Normalized cross-correlation function versus the number of images:} left images: correlation computed on the physical pixels (only the central part is presented). Smaller right images: correlation computed after grouping $8\times8$ pixels .(\textbf{a},\textbf{b}), far-field. (\textbf{c},\textbf{d}), near field}.
\label{SNR}
\end{figure*}
\begin{table}[ht]
\caption{Inferred variances.}
\centering
\begin{tabular}{|l r|}
\hline
Variances & Measured values\\
\hline
$\Delta^2(x_1-x_2)\ \ $ & $\ \ 299\pm14\ \mu m^2$ \\
$\Delta^2(y_1-y_2)\ \ $ & $\ \ 168\pm7\ \mu m^2$ \\
$\Delta^2(p_{x1}-p_{x2})\ \ $ & $\ \ (9.70\pm0.1)\cdot 10^{-6}\hbar^2\ \mu m^{-2}$ \\
$\Delta^2(p_{y1}-p_{y2})\ \ $ & $\ \ (2.53\pm0.04)\cdot 10^{-6}\hbar^2\ \mu m^{-2}$ \\
\hline
\end{tabular}
\label{tab}
\end{table}
Using the  measured values given in Tab. 1, we find the following product of conditional variances:
\begin{eqnarray}
\Delta^2(x_1-x_2)\Delta^2(p_{x1}+p_{x2})&=&(2.9\pm0.2)\cdot 10^{-3}\hbar^2 \\
\Delta^2(y_1-y_2)\Delta^2(p_{y1}+p_{y2})&=&(4.2\pm0.2)\cdot 10^{-4}\hbar^2
\end{eqnarray}
Those results clearly violate inequality (\ref{EPR}), thus exhibiting an EPR paradox in the two transverse dimensions. The variance product of the experimental values is more than 1,200 standard deviations under the classical limit along $x$ and more than 12,000 standard deviations under this limit along $y$. Moreover, the results are in rather good agreement with the theoretical expectations $8.6\cdot10^{-4}\hbar^2$ on $x$ and $2.6\cdot10^{-4}\hbar^2$ on y obtained by a numerical computation that takes into account the effect of the width of the interference filter. This effect shown in Fig.\ref{fig3}e explains the anisotropy in Fig.\ref{fig4}a, i.e an enlargement in the $x$ direction for the large values of $p_{x1}$.
By using Eq.(\ref{K}), we find along $x$ a degree of paradox of $86\pm5$ and along $y$ of  $595\pm40$. To the best of our knowledge, this degree of $595$ is the highest ever reported for an EPR paradox, whatever the considered domain.

We show in Fig. \ref{SNR} that the minimum number of images that allows a safe assessment of the correlation peaks in both spaces is of the order of 20. Indeed a quantum correlation peak is evidenced if it cannot be confounded, with high probability, with random fluctuations of the background noise. Without any a priori assumption on the position of the peak, this is ensured with a confidence of 99\% if the magnitude of the true peak is greater than 4,5 standard deviations, for an image of $64\times64$ pixels obtained by summing the correlations on groups of 8x8 pixels. This grouping is performed in order to adapt the size of the effective pixel to the size of the correlation peak. In Fig. \ref{SNR}, we have defined the signal-to-noise-ratio (SNR) as the magnitude of the correlation peak divided by the standard deviation, after grouping, of the correlation image outside the peak area. The minimum number of images necessary to demonstrate entanglement is only two in the far-field, where deterministic distortions appear to be smaller than in the image plane.

Finally, we have verified that the images exhibit a sub-shot-noise statistics in both the near-field and the far-field: $r_{n}=0.9975\pm0.0004$ and $r_{f}=0.9959\pm0.0003$, where $r$ is defined by :
\begin{equation}
r=\frac{\left\langle \Delta^2(N_1-N_2)\right\rangle}{\left\langle N_1+N_2\right\rangle}
\end{equation}
that is, the variance of the photon number difference $N_1(\boldsymbol{\rho})-N_2(\boldsymbol{\rho})$ (and $N_1(\boldsymbol{p})-N_2(-\boldsymbol{p})$ in far field) normalized to be expressed in shot noise units. These experimental results are under the classical limit 1 respectively by more than 5 and 10 standard deviation, witnessing the quantum, i.e  particle  like, character of the correlations\cite{Blanchet2008}. Note that smaller values of $r$ can be obtained by grouping the pixels \cite{Devaux2012}, in accordance with the fact that the quantum correlation peak extends on several pixels.

To conclude, we have demonstrated a two dimensional EPR paradox in the closest form of its original proposal by recording the behavior of light in couples of twin images. The quantum character of these images has been doubly demonstrated firstly by full-field measurement of a high degree of EPR paradox for both transverse directions and secondly by demonstrating sub-shot noise character in both the near-field and the far-field. Reliable results can be obtained with 20 images, i.e. an acquisition time of 4 seconds and a computation time  that scales also in seconds since cross correlations are computed using  FFT algorithms. This should be compared to days for raster scanning, or hours for compressive-sensing \cite{Howland2013}. Because of the experimental anisotropy, the dimensionality of entanglement, or Schmidt number $K$, can be assessed as the square root of the product of the paradox degrees in each direction: $K=\sqrt{594\times85}=225$.  Such high-dimensionality spatial entanglement has applications in numerous fields of quantum optics, like quantum cryptography \cite{Branciard2012} or quantum computation \cite{Tasca2011}.

\bigskip

\section*{Author Contributions}
\noindent E.L. and F.D. designed the experiment. P.A.M. optimized the experimental set-up and performed the experimental work under the supervision of F.D.  P.A.M. made the data treatment and the numerical modeling, with a participation of E.L. P.A.M. and E.L. wrote the initial manuscript text. All authors discussed the results and substantially contributed to the manuscript.
\section*{Additional information}
\noindent The authors declare no competing financial interests.


\begin{thebibliography}{10}
\expandafter\ifx\csname url\endcsname\relax
  \def\url#1{\texttt{#1}}\fi
\expandafter\ifx\csname urlprefix\endcsname\relax\def\urlprefix{URL }\fi
\providecommand{\bibinfo}[2]{#2}
\providecommand{\eprint}[2][]{\url{#2}}

\bibitem{Howland2013}
\bibinfo{author}{Howland, G.~A.} \& \bibinfo{author}{Howell, J.~C.}
\newblock \bibinfo{title}{Efficient high-dimensional entanglement imaging with
  a compressive-sensing double-pixel camera}.
\newblock \emph{\bibinfo{journal}{Physical review X}}
  \textbf{\bibinfo{volume}{3}}, \bibinfo{pages}{011013} (\bibinfo{year}{2013}).

\bibitem{Dixon2012}
\bibinfo{author}{Dixon, P.~B.}, \bibinfo{author}{Howland, G.~A.},
  \bibinfo{author}{Schneeloch, J.} \& \bibinfo{author}{Howell, J.~C.}
\newblock \bibinfo{title}{Quantum mutual information capacity for
  high-dimensional entangled states}.
\newblock \emph{\bibinfo{journal}{Phys. Rev. Lett.}}
  \textbf{\bibinfo{volume}{108}}, \bibinfo{pages}{143603}
  (\bibinfo{year}{2012}).

\bibitem{Einstein1935}
\bibinfo{author}{Einstein, A.}, \bibinfo{author}{Podolsky, B.} \&
  \bibinfo{author}{Rosen, N.}
\newblock \bibinfo{title}{Can quantum-mechanical description of physical
  reality be considered complete?}
\newblock \emph{\bibinfo{journal}{Phys. Rev.}} \textbf{\bibinfo{volume}{47}},
  \bibinfo{pages}{777--780} (\bibinfo{year}{1935}).

\bibitem{Howell2004}
\bibinfo{author}{Howell, J.~C.}, \bibinfo{author}{Bennink, R.~S.},
  \bibinfo{author}{Bentley, S.~J.} \& \bibinfo{author}{Boyd, R.~W.}
\newblock \bibinfo{title}{Realization of the
  \textsc{E}instein-\textsc{P}odolsky-\textsc{R}osen paradox using momentum-
  and position-entangled photons from spontaneous parametric down conversion}.
\newblock \emph{\bibinfo{journal}{Phys. Rev. Lett.}}
  \textbf{\bibinfo{volume}{92}}, \bibinfo{pages}{210403}
  (\bibinfo{year}{2004}).

\bibitem{Reid2009}
\bibinfo{author}{Reid, M.~D.} \emph{et~al.}
\newblock \bibinfo{title}{\textit{Colloquium} : The
  \textsc{E}instein-\textsc{P}odolsky-\textsc{R}osen paradox: from concepts to
  applications}.
\newblock \emph{\bibinfo{journal}{Rev. Mod. Phys.}}
  \textbf{\bibinfo{volume}{81}}, \bibinfo{pages}{1727--1751}
  (\bibinfo{year}{2009}).

\bibitem{Law2004}
\bibinfo{author}{Law, C.} \& \bibinfo{author}{Eberly, J.}
\newblock \bibinfo{title}{Analysis and interpretation of high transverse
  entanglement in otical parametric down conversion}.
\newblock \emph{\bibinfo{journal}{Phys. Rev. Lett.}}
  \textbf{\bibinfo{volume}{12}}, \bibinfo{pages}{127903}
  (\bibinfo{year}{2004}).

\bibitem{Exter2006}
\bibinfo{author}{van Exter, M.}, \bibinfo{author}{Aiello, A.},
  \bibinfo{author}{Oemrawsingh, S.}, \bibinfo{author}{Nienhuis, G.} \&
  \bibinfo{author}{Woerdman, J.~P.}
\newblock \bibinfo{title}{Effect of spatial filtering on the \textsc{S}chmidt
  decomposition of entangled photons}.
\newblock \emph{\bibinfo{journal}{Phys. Rev. A}} \textbf{\bibinfo{volume}{74}},
  \bibinfo{pages}{012309} (\bibinfo{year}{2006}).

\bibitem{Devaux95}
\bibinfo{author}{Devaux, F.} \& \bibinfo{author}{Lantz, E.}
\newblock \bibinfo{title}{Transfer function of spatial frequencies in
  parametric amplification : experimental analysis and application to
  picosecond spatial filtering}.
\newblock \emph{\bibinfo{journal}{Optics Communications}}
  \textbf{\bibinfo{volume}{114}}, \bibinfo{pages}{295--300}
  (\bibinfo{year}{1995}).

\bibitem{Bell1964}
\bibinfo{author}{Bell, J.~S.}
\newblock \bibinfo{title}{On the
  \textsc{E}instein-\textsc{P}odolsky-\textsc{R}osen paradox}.
\newblock \emph{\bibinfo{journal}{Physics}} \textbf{\bibinfo{volume}{1}},
  \bibinfo{pages}{195--200} (\bibinfo{year}{1964}).

\bibitem{Aspect1981}
\bibinfo{author}{Aspect, A.}, \bibinfo{author}{Grangier, P.} \&
  \bibinfo{author}{Roger, G.}
\newblock \bibinfo{title}{Experimental tests of realistic local theories via
  \textsc{B}ell's theorem}.
\newblock \emph{\bibinfo{journal}{Phys. Rev. Lett.}}
  \textbf{\bibinfo{volume}{47}}, \bibinfo{pages}{460--463}
  (\bibinfo{year}{1981}).

\bibitem{Brida2010}
\bibinfo{author}{Brida, G.}, \bibinfo{author}{Genovese, M.} \&
  \bibinfo{author}{Berchera, I.~R.}
\newblock \bibinfo{title}{Experimental realization of sub-shot-noise quantum
  imaging}.
\newblock \emph{\bibinfo{journal}{Nat Photon}} \textbf{\bibinfo{volume}{4}},
  \bibinfo{pages}{227--230} (\bibinfo{year}{2010}).

\bibitem{Ottavia04}
\bibinfo{author}{Jedrkiewicz, O.} \emph{et~al.}
\newblock \bibinfo{title}{Detection of sub-shot-noise spatial correlation in
  high-gain parametric down conversion}.
\newblock \emph{\bibinfo{journal}{Phys. Rev. Lett.}}
  \textbf{\bibinfo{volume}{93}}, \bibinfo{pages}{243601}
  (\bibinfo{year}{2004}).

\bibitem{Lantz2008}
\bibinfo{author}{Lantz, E.}, \bibinfo{author}{Blanchet, J.-L.},
  \bibinfo{author}{Furfaro, L.} \& \bibinfo{author}{Devaux, F.}
\newblock \bibinfo{title}{Multi-imaging and \textsc{B}ayesian estimation for
  photon counting with \textsc{EMCCD}s}.
\newblock \emph{\bibinfo{journal}{Monthly Notices of the Royal Astronomical
  Society}} \textbf{\bibinfo{volume}{386}}, \bibinfo{pages}{2262--2270}
  (\bibinfo{year}{2008}).

\bibitem{Oemrawsingh2002}
\bibinfo{author}{Oemrawsingh, S. S.~R.}, \bibinfo{author}{van Drunen, W.~J.},
  \bibinfo{author}{Eliel, E.~R.} \& \bibinfo{author}{Woerdman, J.~P.}
\newblock \bibinfo{title}{Two-dimensional wave vector correlations in
  spontaneous parametric downconversion explored with an intensified ccd
  camera}.
\newblock \emph{\bibinfo{journal}{J. Opt. Soc. Am. B}}
  \textbf{\bibinfo{volume}{19}}, \bibinfo{pages}{2391--2395}
  (\bibinfo{year}{2002}).

\bibitem{Fickler2013}
\bibinfo{author}{Fickler, R.}, \bibinfo{author}{Krenn, M.},
  \bibinfo{author}{Lapkiewicz, R.}, \bibinfo{author}{Ramelow, S.} \&
  \bibinfo{author}{Zeilinger, A.}
\newblock \bibinfo{title}{Real-time imaging of quantum entanglement}.
\newblock \emph{\bibinfo{journal}{Sci. Rep.}} \textbf{\bibinfo{volume}{3}},
  \bibinfo{pages}{--} (\bibinfo{year}{2013}).

\bibitem{Blanchet2008}
\bibinfo{author}{Blanchet, J.-L.}, \bibinfo{author}{Devaux, F.},
  \bibinfo{author}{Furfaro, L.} \& \bibinfo{author}{Lantz, E.}
\newblock \bibinfo{title}{Measurement of sub-shot-noise correlations of spatial
  fluctuations in the photon-counting regime}.
\newblock \emph{\bibinfo{journal}{Phys. Rev. Lett.}}
  \textbf{\bibinfo{volume}{101}}, \bibinfo{pages}{233604}
  (\bibinfo{year}{2008}).

\bibitem{Blanchet2010}
\bibinfo{author}{Blanchet, J.-L.}, \bibinfo{author}{Devaux, F.},
  \bibinfo{author}{Furfaro, L.} \& \bibinfo{author}{Lantz, E.}
\newblock \bibinfo{title}{Purely spatial coincidences of twin photons in
  parametric spontaneous down-conversion}.
\newblock \emph{\bibinfo{journal}{Phys. Rev. A}} \textbf{\bibinfo{volume}{81}},
  \bibinfo{pages}{043825} (\bibinfo{year}{2010}).

\bibitem{Moreau2012}
\bibinfo{author}{Moreau, P.-A.}, \bibinfo{author}{Mougin-Sisini, J.},
  \bibinfo{author}{Devaux, F.} \& \bibinfo{author}{Lantz, E.}
\newblock \bibinfo{title}{Realization of the purely spatial
  \textsc{E}instein-\textsc{P}odolsky-\textsc{R}osen paradox in full-field
  images of spontaneous parametric down-conversion}.
\newblock \emph{\bibinfo{journal}{Phys. Rev. A}} \textbf{\bibinfo{volume}{86}},
  \bibinfo{pages}{010101} (\bibinfo{year}{2012}).

\bibitem{Edgar2012}
\bibinfo{author}{Edgar, M.} \emph{et~al.}
\newblock \bibinfo{title}{Imaging high-dimensional spatial entanglement with a
  camera}.
\newblock \emph{\bibinfo{journal}{Nat Commun}} \textbf{\bibinfo{volume}{3}},
  \bibinfo{pages}{984--} (\bibinfo{year}{2012}).

\bibitem{Devaux2012}
\bibinfo{author}{Devaux, F.}, \bibinfo{author}{Mougin-Sisini, J.},
  \bibinfo{author}{Moreau, P.-A.} \& \bibinfo{author}{Lantz, E.}
\newblock \bibinfo{title}{Towards the evidence of a purely spatial
  \textsc{E}instein-\textsc{P}odolsky-\textsc{R}osen paradox in images:
  measurement scheme and first experimental results}.
\newblock \emph{\bibinfo{journal}{The European Physical Journal D}}
  \textbf{\bibinfo{volume}{66}}, \bibinfo{pages}{1--6} (\bibinfo{year}{2012}).

\bibitem{Reid1989}
\bibinfo{author}{Reid, M.~D.}
\newblock \bibinfo{title}{Demonstration of the
  \textsc{E}instein-\textsc{P}odolsky-\textsc{R}osen paradox using
  nondegenerate parametric amplification}.
\newblock \emph{\bibinfo{journal}{Phys. Rev. A}} \textbf{\bibinfo{volume}{40}},
  \bibinfo{pages}{913--923} (\bibinfo{year}{1989}).

\bibitem{Tasca2009}
\bibinfo{author}{Tasca, D.~S.}, \bibinfo{author}{Walborn, S.~P.},
  \bibinfo{author}{Souto~Ribeiro, P.~H.}, \bibinfo{author}{Toscano, F.} \&
  \bibinfo{author}{Pellat-Finet, P.}
\newblock \bibinfo{title}{Propagation of transverse intensity correlations of a
  two-photon state}.
\newblock \emph{\bibinfo{journal}{Phys. Rev. A}} \textbf{\bibinfo{volume}{79}},
  \bibinfo{pages}{033801} (\bibinfo{year}{2009}).

\bibitem{Fedorov2009}
\bibinfo{author}{Fedorov, M.}, \bibinfo{author}{Mikhailova, Y.~M.} \&
  \bibinfo{author}{Volkov, P.~A.}
\newblock \bibinfo{title}{Gaussian modelling and \textsc{S}chmidt modes of
  \textsc{SPDC} biphoton states}.
\newblock \emph{\bibinfo{journal}{J. Phys. B}} \textbf{\bibinfo{volume}{42}},
  \bibinfo{pages}{175503} (\bibinfo{year}{2009}).

\bibitem{Lantz2000}
\bibinfo{author}{Lantz, E.} \& \bibinfo{author}{Devaux, F.}
\newblock \bibinfo{title}{The phase-mismatch vector and resolution in image
  parametric amplification}.
\newblock \emph{\bibinfo{journal}{Journal of Optics A}}
  \textbf{\bibinfo{volume}{2}}, \bibinfo{pages}{362--364}
  (\bibinfo{year}{2000}).

\bibitem{Branciard2012}
\bibinfo{author}{Branciard, C.}, \bibinfo{author}{Cavalcanti, E.~G.},
  \bibinfo{author}{Walborn, S.~P.}, \bibinfo{author}{Scarani, V.} \&
  \bibinfo{author}{Wiseman, H.~M.}
\newblock \bibinfo{title}{One-sided device-independent quantum key
  distribution: Security, feasibility, and the connection with steering}.
\newblock \emph{\bibinfo{journal}{Phys. Rev. A}} \textbf{\bibinfo{volume}{85}},
  \bibinfo{pages}{010301} (\bibinfo{year}{2012}).

\bibitem{Tasca2011}
\bibinfo{author}{Tasca, D.~S.}, \bibinfo{author}{Gomes, R.~M.},
  \bibinfo{author}{Toscano, F.}, \bibinfo{author}{Souto~Ribeiro, P.~H.} \&
  \bibinfo{author}{Walborn, S.~P.}
\newblock \bibinfo{title}{Continuous-variable quantum computation with spatial
  degrees of freedom of photons}.
\newblock \emph{\bibinfo{journal}{Phys. Rev. A}} \textbf{\bibinfo{volume}{83}},
  \bibinfo{pages}{052325} (\bibinfo{year}{2011}).

\end{thebibliography}
\end{document}